\documentstyle[sprocl,epsfig]{article}

\def\EPJ{{\em Eur. Phys. J.} C }

\def\be{\begin{equation}}
\def\ee{\end{equation}}
\def\bea{\begin{eqnarray}}
\def\eea{\end{eqnarray}}

\newcommand{\m}{\tiny{-}}
\newcommand{\p}{\tiny{+}}

\def\beq{\begin{equation}}
\def\eeq{\end{equation}}
\def\beqar{\begin{eqnarray}}
\def\eeqar{\end{eqnarray}}
\def\barr#1{\begin{array}{#1}}
\def\earr{\end{array}}
\def\bfi{\begin{figure}}
\def\efi{\end{figure}}
\def\btab{\begin{table}}
\def\etab{\end{table}}
\def\bce{\begin{center}}
\def\ece{\end{center}}

\def\text{\textstyle}

\def\be{\beta}

\def\mathswitchr#1{\relax\ifmmode{\mathrm{#1}}\else$\mathrm{#1}$\fi}

\def\mathswitch#1{\relax\ifmmode#1\else$#1$\fi}

\def\stars{\strut\leaders\hbox{*}\hfill\strut}
\def\starline{\hfil\strut\hfil\hbox to \textwidth {\stars}\hfil}

\newcommand{\bc}{\begin{center}}
\newcommand{\ec}{\end{center}}

\newcommand{\Dir}{\kern -6.4pt\Big{/}}
\newcommand{\Dirin}{\kern -10.4pt\Big{/}\kern 4.4pt}
\newcommand{\DDir}{\kern -8.0pt\Big{/}}
\newcommand{\DGir}{\kern -6.0pt\Big{/}}

\def\slashchar#1{\setbox0=\hbox{$#1$}           
     \dimen0=\wd0                                 
     \setbox1=\hbox{/} \dimen1=\wd1               
     \ifdim\dimen0>\dimen1                        
        \rlap{\hbox to \dimen0{\hfil/\hfil}}      
        #1                                        
     \else                                        
        \rlap{\hbox to \dimen1{\hfil$#1$\hfil}}   
        /                                         
     \fi}                                         %

\def\be{\begin{equation}}
\def\ee{\end{equation}}
\def\bea{\begin{eqnarray}}
\def\eea{\end{eqnarray}}

\def\gsim{\:\raisebox{-0.5ex}{$\stackrel{\textstyle>}{\sim}$}\:}

\def\slash{/\kern -5pt}

\def\ims #1 {\ensuremath{M^2_{[#1]}}}

\def\s22w{s_{2W}^2}

\begin{document}
\noindent
SHEP-04-24
\hskip7.cm
LC-TOOL-2004-018

\title{SIX-FERMIONS (AND MORE) STUDIES\\}

\author{ STEFANO MORETTI }

\address{School of Physics and Astronomy, University of Southampton,\\
Highfield, Southampton SO17 1BJ, UK}


\maketitle\abstracts{
We review the available event generators 
suited for multi-fermion final state production in the context of physics
studies at a future Linear Collider (LC)   
}

\section{Why six (and more !) fermions at LCs ?}

While LEP1 (${\sqrt s= M_Z}$)  
was the realm of two-fermion processes ($e^+e^-\to Z\to f\bar f$,
$f=q,\ell$) and LEP2 ($\sqrt s\gsim 2 M_V$, $V=W,Z$)
was the arena for four-fermion reactions
($e^+e^-\to VV\to f\bar f f'\bar f'$), future
 LCs ($\sqrt s =350-800$ {\rm{GeV}})
will open the era of multi-fermion channels in both
Standard Model (SM), 
\begin{eqnarray} \nonumber  
e^+e^- & \to & t\bar t  \to  (bW^+)(\bar bW^-)\to 6f~~({\rm{Top-quarks}}),\\ \nonumber 
       & \to & Zh \to  (f\bar f)(VV)\to 6f~~({\rm{Higgs-strahlung}}),\\ \nonumber
       & \to & \nu_e\bar\nu_e[e^+e^-] h \to \nu_e\bar\nu_e[e^+e^-]  (VV)
\to 6f~~({\rm{Vector~Boson~Fusion}}),\\ \nonumber
       & \to & W^+W^-Z[ZZZ] \to  \to 6f~~({\rm{Quartic~Gauge~Couplings}}),\\ \nonumber
       & \to & Zhh \to  (f\bar f)(b\bar b)(b\bar b)~~({\rm{Higgs~self-couplings}}),\\ \nonumber
       & \to & \nu_e\bar\nu_e[e^+e^-] hh \to \nu_e\bar\nu_e[e^+e^-]  (b\bar b)(b\bar b)~~({\sl{ditto}}),\\ \nonumber
       & \to & t\bar t  h \to  8{{f}}~~({\rm{Top-Yukawa-coupling}}),\\ \nonumber 
\end{eqnarray}
\vskip-0.5cm\noindent
and general 2-Higgs Doublet Models, 
\begin{eqnarray} \nonumber
e^+e^-  \to   AH &\to&  (b\bar b) (VV) \to 6{{f}}~~({\rm{Pseudoscalar-Higgs}}), \\ \nonumber   
e^+e^-       \to  H^+H^-  &\to&  (t\bar b)
(\tau^-\bar\nu_\tau) [(t\bar b)  (\bar tb)]\to 6{{f}}[8f]~~({\rm{Charged-Higgs}}).
\end{eqnarray}
Given the interesting physics accessible through
multi-fermion processes, it is of paramount importance to {\sl develop} suitable computational tools as well as {\sl test} these in view of their phenomenological
applications.

\section{Two categories of computational tools}

$1.$ Standard Monte Carlo (MC) {\sl{event}} generators: e.g., 
{\tt PYTHIA} \cite{pythia}, 
{\tt HERWIG} \cite{herwig}
and 
{\tt ISAJET} \cite{isajet}, wherein the above final states are typically 
produced via resonant subprocesses ({e.g.}, $t\bar t$, $H^+H^-$, etc.) and
with no irreducible background but have
added QCD (and QED) Parton Shower (PS) and hadronisation. 
\vskip0.15cm\noindent
$2.$ So-called {\sl{parton}} generators: they can 
compute both signal and irreducible background in multi-fermion
processes (including interference effects) but traditionally lack 
the PS and hadronisation treatment of quarks and gluons.
The latter can be further categorised in:
\vskip0.15cm\noindent
{\underbar{{Multi-purpose generators,}}} such as:
{\tt GRACE} \cite{grace};
{\tt HELAS/MadGraph/MadEvent} \cite{helas,madgraph};
{\tt Whi\-zard+Omega/MadGraph/CompHEP} \cite{whizard,omega,comphep};
{\tt SHERPA/AMEGIC++} \cite{sherpa};
{\tt HELAC/PHEGAS} \cite{helac,phegas}.
Their appealing features are that they can provide 
in a (semi-)automated way a vast choice of
final states, with a uniform setup for different process classes, thus being
ideally suited for studying the physics potential of future colliders. Their
drawbacks are that they are lengthy codes, in which the insertion 
of radiative corrections becomes problematic, so that they
are not  normally used as high-precision tools.
\vskip0.15cm\noindent
{\underbar{{Dedicated generators ($e^+e^-\to6f$, no $8f$ yet)}}}, such as:
{\tt eett6f} \cite{eett6f}, for $e^+e^-\to t\bar t\to b\bar b+4f$
processes, including QCD graphs;
{\tt Lusifer} \cite{lusifer}, generating all $6f$ final states
(with massless fermions), but not via ${\cal O}(\alpha_s^4)$;
{\tt SIXFAP} \cite{sixfap}, wherein all $6f$ final states are available (with
massive fermions) but the inclusion of QCD effects is still in progress. 
(A similar code,
currently under development, is {\tt SIXPHACT} \cite{sixphact}.)
The pluses of these codes is that they are optimised for efficient
event generation and that higher order effects
can easily be included, thus rendering them ideal tools 
for high precision studies. The
minuses are that the number of final states is necessarily
limited. 

\vskip0.15cm\noindent
Finally, it should be noted that for $f=q$, where $q$ represent a
quark, only {\sl{jets}} can be observed, so that gluon final states need to be included too: {e.g.},
{$q\bar q q'\bar q' gg$} and
{$q\bar q gggg$} alongside {$q\bar q q'\bar q' q''\bar q'' $},
through 
${\cal O}(\alpha_s^4)$. 
Multi-purpose generators
can compute these gluon final states and also 
dedicated codes (such as {{\tt{SIXRAD}}} \cite{sixrad}) 
exist. Besides, an interface to 
the PS is required for their phenomenological investigations, e.g.,
through the Les Houches Accord (LHA) \cite{LHA}. ({{\tt{SHERPA}}} has 
its own PS, via the {{\tt{APACIC++}}} routines  \cite{apacic}.)

\section{Current studies}

The `Generators' Working Group (WG) of the 
current ECFA Study of Physics and Detectors for a Linear Collider
is engaged in 
systematic studies of multi-fermion final states
following a three-step procedure: 1) test of matrix elements and phase space;
2) comparisons for physics-oriented observables; 3) detector studies.
(Comparisons can be automated by using 
MC-tester  \cite{MCtester} and Java interfaces  \cite{Java}.)
Stage 1) has now been completed.  In Tab.~1
we present a sample of the many results obtained so far:
 see \cite{Stefan,Frank-Costas} and  \cite{Amsterdam} for setup.
The codes display a remarkable agreement 
thus motivating  the WG to moving onto phase 2).
\vskip0.25cm\hskip-1.10cm
$\begin{array}{lcccccc}
\hline
{\footnotesize \sigma~[{\rm{fb}}]} & 
{\mbox{\normalsize\tt AMEGIC++}} & 
{\mbox{\normalsize\tt eett6f}} & 
{\mbox{\normalsize\tt Lusifer}} & 
{\mbox{\normalsize\tt HELAC}} & 
{\mbox{\normalsize\tt SIXFAP}} & 
{\mbox{\normalsize\tt Whizard}} \\
\hline
\nu_ee^+e^-\bar\nu_e b\bar b &
	5.879(8) & 5.862(6)         & 5.853(7) & 5.866(9) & 5.854(3) & 5.875(3) \\
\nu_ee^+\mu^-\bar\nu_\mu b\bar b &
	5.827(4) & 5.815(5)  & 5.819(5) & 5.822(7) & 5.815(2) & 5.827(3) \\
\nu_\mu\mu^+\mu^-\bar\nu_\mu b\bar b &
	5.809(5) & 5.807(3) & 5.809(5) & 5.809(5) & 5.804(2) & 5.810(3) \\
\nu_\mu\mu^+\tau^-\bar\nu_\tau b\bar b &
	5.800(3) & 5.797(5) & 5.800(4) & 5.798(4) & 5.798(2) & 5.796(3) \\
\hline
\end{array}$ 
\vspace*{-0.05truecm}
\begin{center}
\begin{tabular}{lccc} 
$\sigma$  [fb]              & QCD & {\tt AMEGIC++}&
{\tt HELAC}\\[-.3em]
\noalign{\smallskip}\hline\noalign{\smallskip}
$\mu^{\m}\mu^{\p}b \bar bb \bar b$  
                     & yes & 3.096(60)e-02 &3.019(43)e-02\\
                     & no  & 2.34(12)e-02   &2.36(10)e-02\\[-.3em]
\noalign{\smallskip}\hline
\end{tabular}
\end{center}
{Table 1: Comparison among multi-fermion codes for selected final states.}

\end{document}